\documentclass[twocolumn,footinbib,notitlepage,showpacs,amsmath,amssymb,pra,aps,10pt]{revtex4-1}
\usepackage{units}
\usepackage[pdftex]{graphicx}              
\usepackage[pdftex]{hyperref}
\usepackage{pdfsync}

\begin{document}

\title{Time-averaged adiabatic ring potential for ultracold atoms}

\author{B. E. Sherlock}
\author{M. Gildemeister}
\author{E. Owen}
\author{E. Nugent}
\author{C. J. Foot}
\affiliation{Clarendon Laboratory, University of Oxford, Parks Road, Oxford, OX1 3PU, United Kingdom}

\date{\today}

\begin{abstract}
We report the experimental realisation of a versatile ring trap for ultracold atoms. The ring geometry is created by the time-averaged adiabatic potential resulting from the application of an oscillating magnetic bias field to a rf dressed quadrupole trap. Lifetimes for a Bose-Einstein condensate in the ring exceed \unit[11]{s} and the ring radius was continuously varied from $\unit[50]{\mu \textrm{m}}$ to $\unit[262]{\mu\textrm{m}}$. An efficient method of loading the ring from a conventional TOP trap is presented together with a rotation scheme which introduces angular momentum into the system. The ring presents an opportunity to study the superfluid properties of a condensate in a multiply connected geometry and also has applications for matter-wave interferometry.  
\end{abstract}


\maketitle

\section{Introduction}
\label{sec:intro}
The development of increasingly sophisticated trapping potentials for ultracold atoms has catalysed research in the field of quantum degenerate gases. Optical lattices \cite{greiner}, double well potentials \cite{albiez} and systems of reduced dimensionality \cite{paredes} are examples where new physics has been elucidated by putting cold atoms into new types of potential landscapes. The geometry of ring-shaped traps gives rise to multiply connected systems with behaviour not found in other types of trap. 

For a ring trap there are two regimes that can be distinguished. In the first regime the quantum coherence extends all around the ring. Under these conditions the superfluid nature of an interacting dilute ultracold quantum gas is manifested, e.g. persistent flow. The angular momentum of a superfluid confined in annular geometry dissipates only if the rotational velocity exceeds a critical value \cite{raman, neely}. Below the critical velocity the mass flow is dissipationless in analogy to electrical current in superconductors. Such a persistent flow was observed with ultracold atoms in an optically plugged magnetic trap \cite{ringtrapPhillips}. Recently an all optical ring trap experiment demonstrated that superflow around a ring is suppressed by a barrier where the local flow velocity exceeds the critical value \cite{ringtrapPhillips2}. This may lead to an analogue of the superconducting quantum interference device (SQUID) using a Bose-Einstein condensate (BEC).

The second regime employs the ring trap as a wave guide for matter-wave interferometry. In combination with coherent beam-splitting, a circular waveguide forms a Sagnac interferometer \cite{sagnac} and acts as an inertial sensing device. Several experiments have been designed with this goal in mind \cite{FirstAtomRing2001, GuptaRing, ArnoldLargeRing}, and further schemes have been proposed \cite{morizot_ring, ArnoldInductiveRing, baker}.

The two regimes can be loosely distinguished by the constraints placed on the chemical potential for a BEC in 3D ring trap \cite{morizot_ring}:
\begin{equation}
\mu_{3D} = \hbar \bar{\omega}\sqrt{\frac{2 a_s N}{\pi r_0}}.
\label{eq:chemPot}
\end{equation}
Here $\bar{\omega}=\sqrt{\omega_z \omega_r}$ is the geometric mean of the axial ($\omega_{z}$) and radial ($\omega_{r}$) trapping frequency, $a_s$ the $s$-wave scattering length, $N$ the atoms number and $r_0$ the ring radius. For the first regime, $\mu$ must be large compared to the potential inhomogeneities around the ring, enabling the BEC to spread around the full ring circumference. In the second regime the condensate is localised by a variation in the potential around the ring. 
\begin{figure}[t]
\begin{center}
\includegraphics[width=1.0\columnwidth]{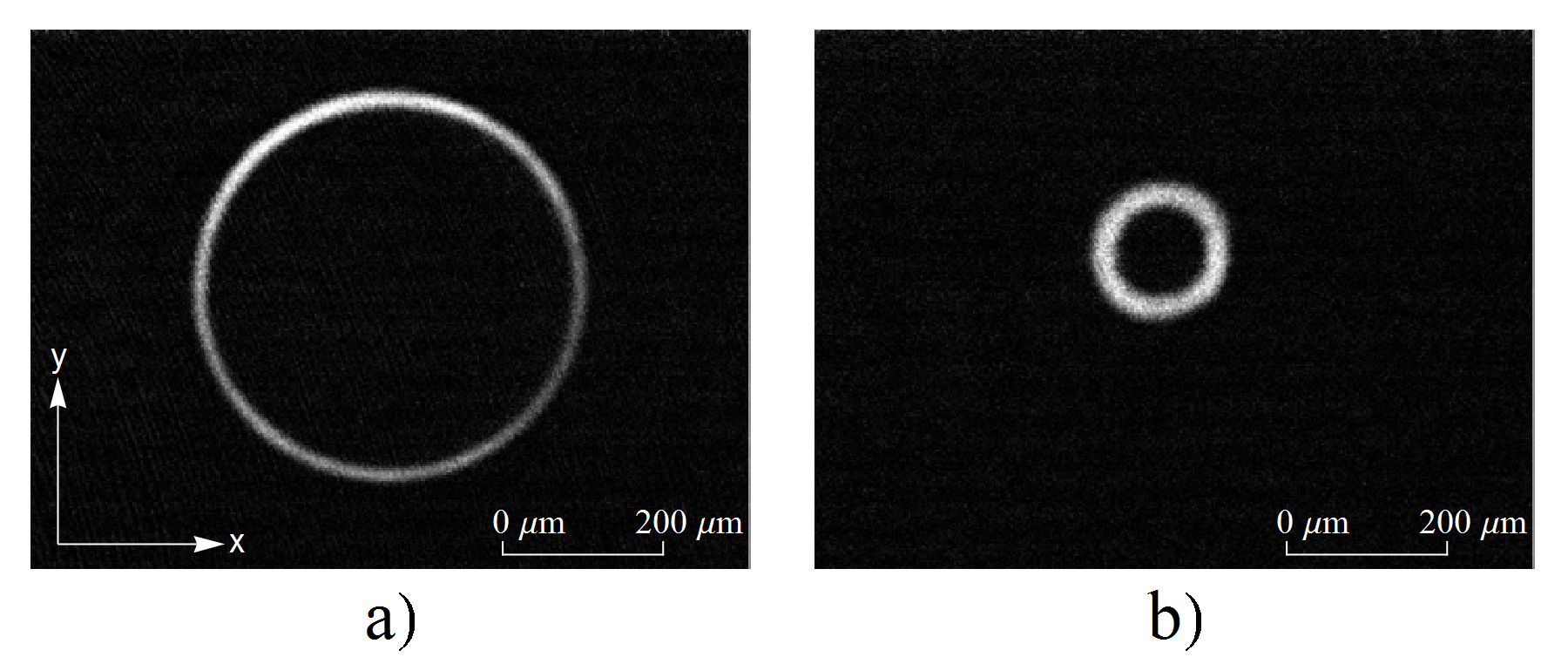}
\caption{Absorption image of atoms confined in the TAAP ring trap. In (a) the cloud of atoms has a radius of $\unit[262]{\mu \textrm{m}}$ and in (b) the radius is $\unit[71]{\mu \textrm{m}}$. This change is achieved by increasing the gradient of the quadrupole field from (a) $\unit[65.3]{\textrm{G/cm}}$ to (b) $\unit[277.1]{\textrm{G/cm}}$. In both cases $\omega_{rf} = 2\pi \times \unit[1.4]{\textrm{MHz}}$, $B_{rf} = \unit[0.8]{\textrm{G}}$ and $B^{z}_{T}= \unit[0.9]{\textrm{G}}$.
\label{fig:rings}}
\end{center}
\end{figure}

The ring trap described here aspires to both regimes. It is based on the proposal for a time-averaged adiabatic potential (TAAP) ring trap \cite{Lesanovsky2007}. In this particular experiment a rf dressed quadrupole trap is time-averaged along its symmetry axis which results in a ring-shaped trap geometry (see figure \ref{fig:rings}) with a range of radii suitable for exploring both of the above regimes.

In previous work a rf dressed potential was combined with two blue frequency detuned light sheets to create a ring potential \cite{heathcote}. This arrangement was susceptible to thermally induced drifts in the position of the optical potential in which atoms had only moderate lifetimes ($\tau \sim \unit[1]{\textrm{s}}$). The new experimental design depends only on magnetic potentials which do not require repeated alignment. Increased power in the rf fields has resulted in lifetimes $> \unit[11]{\textrm{s}}$, whilst a custom designed frequency source has led to improved dynamical control over the rf polarisation allowing the trap to be tilted and rotated.   

This article is organized as follows: in section \ref{sec:theory} the theory of the time-averaged adiabatic potential is discussed and the formation of the ring is explained. Section \ref{sec:results} outlines the experimental sequence and details the ring trap characterisation results and their comparison with numerical simulations. Section \ref{sec:rotation} describes the rotation scheme that has been implemented in the ring potential and some preliminary results. The article concludes with an outlook in section \ref{sec:outlook}.

\section {Theory}
\label{sec:theory}
The magnetostatic potential experienced by an atom in an inhomogeneous magnetic field is modified by the application of strong rf radiation, in the regions where the radiation drives transitions between internal states of the atom. Zeeman sublevels of the atomic hyperfine ground state are magnetically coupled together via an external field oscillating at a frequency $\omega_{rf}$ comparable to the local Larmor frequency $\omega_{0}(\textrm{\textbf{r}}) = \left|\textrm{g}_{F} \mu_{B}\textbf{B}(\textrm{\textbf{r}})/\hbar\right|$ where $\textrm{\textbf{B}}(\textrm{\textbf{r}})$ is the local static field, $\textrm{g}_{F}$ the Land\'{e} g-factor and $\mu_{B}$ the Bohr magneton. 

Following the dressed atom picture \cite{haroche} the eigenstates of the global (atom $+$ field mode $+$ interaction) Hamiltonian, known as dressed states, consist of spatially dependent superpositions of the bare atomic energy levels. In the limit of strong coupling the dressed states form an adiabatic potential (AP) described by,
\begin{equation}
U_{AP}(\textrm{\textbf{r}})=m_{F} \hbar \sqrt{\delta^2(\textrm{\textbf{r}})+\Omega^2_{R}(\textrm{\textbf{r}})}
\label{AP}
\end{equation}
where $\delta(\textrm{\textbf{r}})=\omega_{rf}-\omega_{0}(\textrm{\textbf{r}})$ is the angular frequency detuning and $\Omega_{R}(\textrm{\textbf{r}})$ is the Rabi coupling frequency between the bare states. At $\delta(\textrm{\textbf{r}})=0$ the resonance condition $\left|\textrm{g}_{F} \mu_{B} \textrm{\textbf{B}}(\textrm{\textbf{r}})/\hbar\right| = \omega_{rf}$ is satisfied. Neglecting the coupling term $\Omega_{R}(\textrm{\textbf{r}})$, it is apparent that the locus of minimum energy lies on an isosurface of constant $\left|\textbf{\textrm{B}}(\textbf{\textrm{r}})\right|$ defined by the resonance condition. 

The Rabi frequency term
\begin{equation}
\Omega_{R}(\textrm{\textbf{r}}) = \left|\frac{\textrm{g}_{F} \mu_{B}}{2\hbar}\frac{\textrm{\textbf{B}}(\textrm{\textbf{r}})}{\left|\textrm{\textbf{B}}(\textrm{\textbf{r}})\right|}\times \textbf{\textrm{B}}_{rf}\right|,
\label{coupling}
\end{equation}
has a vectorial nature that derives from the requirement that transitions between neighbouring $m_{F}$ states are only driven by a field oscillating perpendicular to the orientation of the magnetic dipole. In a spatially inhomogeneous magnetic field, $\Omega_{R}$ takes a maximum value when the rf polarisation vector is perpendicular to the local magnetic field vector, and correspondingly a zero value where these fields are collinear. 

\begin{figure}
\begin{center}
\includegraphics[width=0.8\columnwidth]{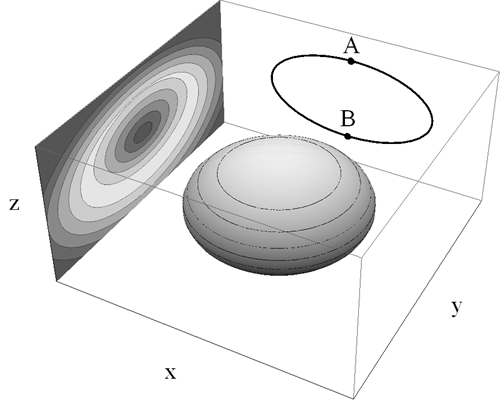}
\caption{The isosurface of constant $\left|\textbf{\textrm{B}}(\textbf{\textrm{r}})\right|$ for a rf dressed quadrupole on which the atoms are trapped. The atoms' position around the ellipsoid is determined by the balance between gravity and the variation of the Rabi frequency $\Omega_{R}(\textbf{\textrm{r}})$. A cross section of the potential (with gravity included) in the $x=0$ plane for circularly polarised rf is projected on the left side of the figure. For this polarisation $\Omega_{R}(\textbf{\textrm{r}})$ varies smoothly from zero at A to its maximum value at B where the ellipse is the projection of the isosurface.  
\label{fig:dressedquadrupole}}
\end{center}
\end{figure}

Strong coupling between bare states induces an avoided crossing between the dressed levels that changes the shape of the potential in the region where $\delta(\textrm{\textbf{r}})\rightarrow0$ and also limits the rate at which trapped atoms make diabatic transitions between dressed states; such Landau Zener (LZ) transitions to untrapped dressed states result in atoms being rapidly ejected from the system. Thus atoms that venture into regions in a potential where $\Omega_{R}(\textrm{\textbf{r}})\rightarrow0$ have vanishingly short lifetimes.

The data presented in this article was taken using a cylindrically symmetric quadrupole field with its symmetry axis aligned with gravity

\begin{equation}
\textrm{\textbf{B}}(\textrm{\textbf{r}})=B'_{q} (x\hat{\textrm{\textbf{e}}}_{x}+y\hat{\textrm{\textbf{e}}}_{y}-2z\hat{\textrm{\textbf{e}}}_{z}).
\label{bquad}
\end{equation}

\begin{figure*}
\includegraphics[width=2\columnwidth]{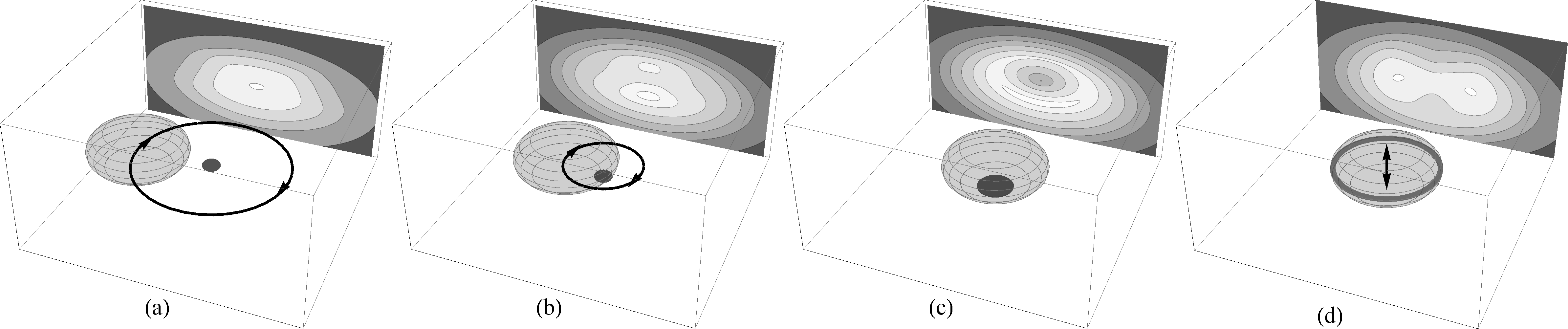}
\caption{The ring trap loading scheme. The bold line shows the locus of the centre of the quadrupole field. Cross sections of the potential the plane $x=0$ are projected at the rear of each plot. (a) A conventional TOP trap. Switching on the rf does not significantly modify the potential experienced by the atoms (represented by the central darkened region) as the resonant ellipsoid orbits exterior to their position. (b) A double-well TAAP trap \cite{gildemeister}. Lowering $B_{T}$ causes the resonant ellipsoid to spiral inwards, trapping the atoms at the two positions where it intersects with the rotation axis. Note due to gravity, in this case only the lower well is loaded. (c) The shell trap. In the absence of a time-averaging field, the atoms are trapped on the resonant ellipsoid, their final position being determined by the balance between gravity and the variation in $\Omega_{R}(\textbf{\textrm{r}})$. (d) The ring trap. Ramping on $B^{z}_{T}$ time-averages the shell potential along its (vertical) symmetry axis. Atoms congregate at the equator of the ellipse (or just below because of gravity).
\label{fig1}
}
\end{figure*}

For this field, the isosurfaces where the resonance condition is satisfied are oblate spheroids centred at the origin $(x=y=z=0)$ where $\left|\textbf{B}(\textrm{\textbf{r}})\right| = 0$. The AP thus takes on an ellipsoidal shell-like geometry as depicted in figure \ref{fig:dressedquadrupole}.

To determine the contribution to the potential landscape of the coupling term $\Omega_{R}(\textrm{\textbf{r}})$ it is necessary to know for a given polarisation, the magnitude of the component of dressing radiation perpendicular to the static field at all positions around the resonant ellipsoid. Analytical solutions for a static quadrupole field and arbitrary rf polarisation are presented in \cite{heathcote}. The ring traps presented in this article were mainly formed using circularly polarised dressing fields of the form, 
\begin{equation}
\textrm{\textbf{B}}_{rf}(t)=B^{x}_{rf}[\cos{(\omega_{rf}t)}\hat{\textrm{\textbf{e}}}_{x}]+B^{y}_{rf}[\sin{(\omega_{rf}t)}\hat{\textrm{\textbf{e}}}_{y}],
\label{circrf}
\end{equation}
where $B^{x}_{rf}=B^{y}_{rf}=B_{rf}$ which results in maximal coupling at the South pole of the ellipsoid. As represented in figure \ref{fig:dressedquadrupole} the coupling strength decreases smoothly around the ellipsoid, reaching zero at the North pole.

Time-averaging of trapping potentials is a versatile technique that has been successfully employed in the electric, optical and magnetic regimes \cite{paul, henderson, petrich}. The process exploits the different timescales that govern an atom's ability to change its motional state, limited by the trapping frequency $\omega_{r}$, and its internal state, closely related to the Larmor frequency $\omega_{0}$. The introduction of an oscillatory magnetic bias field of the form,
\begin{equation}
\textbf{B}_{T}(t)=B_{T}[\cos{(\omega_{T}t)}\hat{\textrm{\textbf{e}}}_{x} + \sin{(\omega_{T}t)}\hat{\textrm{\textbf{e}}}_{y}]
\end{equation}
with a frequency $\omega_{T}$ that lies in the interval $\omega_{r}<\omega_{T}<\omega_{0}$ will result in a trapped atom experiencing a modified potential while preserving its initial $m_{F}$ state.

The realisation of a TAAP trap, created by the application of a time-averaging field to an AP has diversified the trap geometries available. The inherent flexibility of the processes of time-averaging and rf dressing combine to afford access to a broad palette of trap shapes, each of which can be adiabatically loaded and dynamically adjusted to suit experimental requirements \cite{Lesanovsky2007}.

The ring trap described here is a TAAP that results from the application of an axial time-averaging field to a dressed quadrupole trap. When the elliptical geometry of the shell trap is time-averaged along its symmetry (z) axis, the radial curvature of the resulting potential decreases as the  amplitude of the time-averaging field grows. This process continues until zero curvature exists at the initial position of the atoms. A further increase in the amplitude of the axial time-averaging field results in a negative curvature at this position (the bottom of the ellipsoid) and an annular minimum develops giving a ring trap.  

\section{Implementation and Results}
\label{sec:results}

A detailed discussion of the TAAP loading scheme is presented in \cite{gildemeister}. Briefly, a typical experimental sequence begins with \unit[$5 \times 10^{6}$]{atoms} of $^{87}\textrm{Rb}$ at \unit[1]{$\mu$K} confined in the $\left|F=1,m_{F} =-1\right>$ hyperfine state, in an axially symmetric TOP trap ($\omega_{T}=2 \pi \times \unit[7]{kHz}$) where $B_{T} = \unit[3.2]{G}$ and $B'_{q} = \unit[84]{G/cm}$. A rf dressing field (as in equation (\ref{circrf}) with $\omega_{rf}=2 \pi \times \unit[1.4]{MHz}$ and $B_{rf} = \unit[0.8]{G}$ is switched on (see figure 3a). The amplitude of the rotating bias field $B_{T}$ is ramped down over \unit[100]{ms} to a value of \unit[1]{G} during which time the resonant ellipsoid impinges on the position of the atoms and the vertically offset double-well TAAP trap is loaded (see figure 3b). In this potential evaporative cooling for a period of \unit[3]{s} produces a BEC of up to \unit[$4 \times 10^{5}$]{atoms} with no discernible thermal component. The rotating bias field is subsequently ramped to zero over \unit[400]{ms} and the condensate is loaded into the anisotropic shell trap (figure 3c) with oscillation frequencies $\omega_{r}= 2\pi \times$\unit[10]{Hz}, $\omega_{z}= 2\pi \times$\unit[122]{Hz}. The introduction over a period of \unit[1]{s} of an axially directed oscillating bias field $\textbf{B}^{z}_{T}(t)= B^{z}_{T}\cos{(\omega_{T}t)}\hat{\textrm{\textbf{e}}}_{z}$ where $B^{z}_{T} > \unit[0.4]{G}$ changes the shape of the potential minimum into a ring. Provided these manipulations are performed sufficiently slowly to be almost adiabatic, only a small amount of heating occurs and a BEC with a small thermal fraction is loaded into the ring TAAP.

\begin{figure}[h]
\includegraphics[width=0.9\columnwidth]{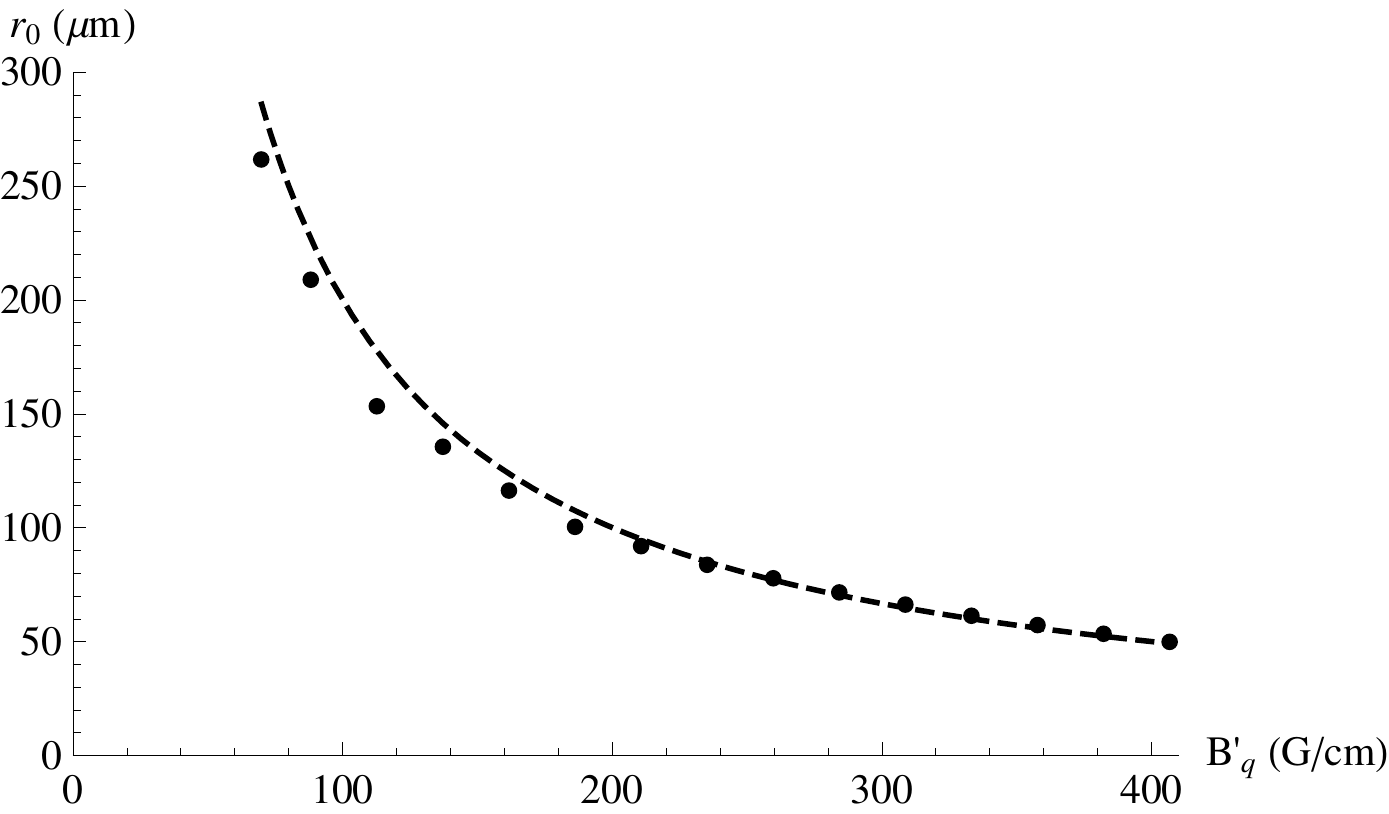}
\caption{Variation of ring radius $r_{0}$ with $B'_{q}$, for $B_{rf} = \unit[0.8]{\textrm{G}}$ and $B^{z}_{T} = \unit[0.9]{\textrm{G}}$. The dashed line represents the theoretical value for semi-major axis of the resonant ellipsoid $r_{e}$ using $\omega_{rf}=\unit[1.4]{\textrm{MHz}}$.
\label{fig:ringradius}
}
\end{figure}

The ring radius $r_{0}$ is determined by $B'_{q}$ and $\omega_{rf}$ which combine through the resonance condition to dictate the length of semimajor axis of the resonant ellipsoid $r_{e}$, i.e.
\begin{equation}
r_{e}=\hbar \omega_{rf}/\left|g_{F} \mu_{B} B'_{q}\right|.
\label{rescon}
\end{equation}
In this experiment the dressing frequency is held constant (at $\omega_{rf} = \unit[1.4]{MHz}$) in order to circumvent practical problems relating to atom heating during rf ramps \cite{morizot_heating}. Despite this condition, as depicted in figure \ref{fig:ringradius}, $r_{0}$ can be continuously varied by adjusting the value of the quadrupole gradient, with the upper limit set by the minimum value of $B'_{q}$ required to support the atoms against gravity. Operating the magnetic trap at the maximum value for $B'_{q} = \unit[410]{\textrm{G/cm}}$ gives a lower limit for $r_{0}$ of \unit[50]{$\mu\textrm{m}$}. There is a weak dependence of $r_{0}$ with $B^{z}_{T}$, however it is typically on the order of a few percent across the full range of oscillatory field amplitudes $\unit[0.4]{G} \leq B^{z}_{T} \leq \unit[1.6]{G} $. The principal result of changing the magnitude of the time-averaging field is to adjust the trap frequencies.

\begin{figure}[ht]
\includegraphics[width=0.8\columnwidth]{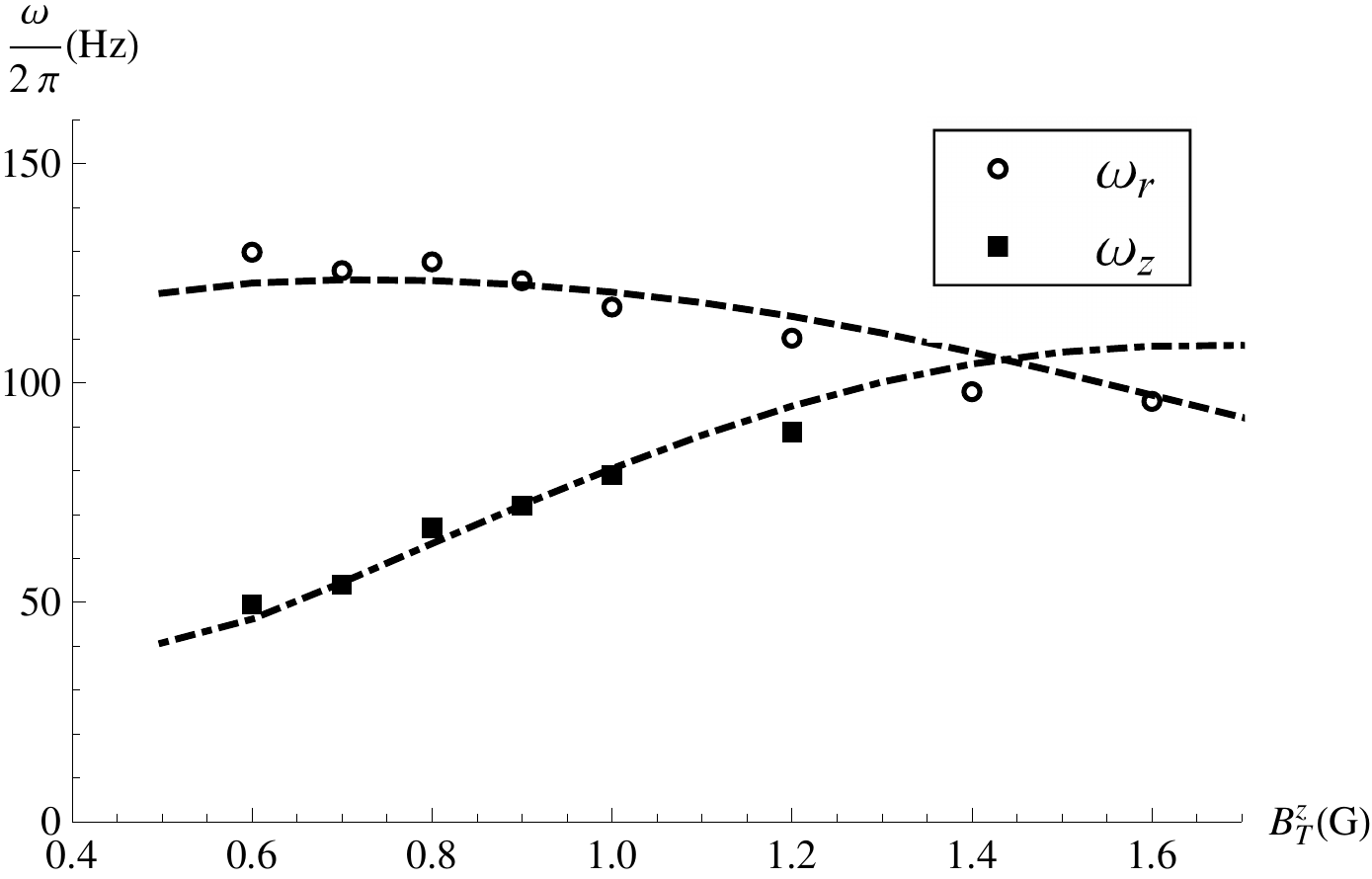}
\caption{Ring trap frequency measurements as a function of $B^{z}_{T}$ with $B'_{q}= \unit[83.5]{G/cm}$ and $B_{rf} = \unit[0.8]{\textrm{G}}$. The dashed and dot-dashed lines represent numerical results plotted for the same experimental parameters. 
\label{fig:trapfreq}
}
\end{figure}

The axial and radial ring trap frequencies were measured against $B^{z}_{T}$ by fitting to dipole mode oscillations in the trapped cloud. Axially oscillations were excited by rapidly displacing the position of the potential minimum using a magnetic bias field. Radially the height of the centrally repulsive region was reduced for a period of \unit[5]{ms} causing the ring radius to oscillate harmonically. The values depicted in figure \ref{fig:trapfreq} compare favourably with numerical simulations and demonstrate how the aspect ratio in the ring can be adjusted. Further results of the numerical analysis suggest that both $\omega_{r}$ and $\omega_{z}$ scale linearly with $B'_{q}$ implying a factor of 9 increase in trap frequencies is readily attainable with the current apparatus.

Lifetimes above \unit[11]{s} are observed in the ring trap when operating at $r_{0}> \unit[200]{\mu \textrm{m}}$. Smaller rings that offer tighter confinement suffer from shorter lifetimes due to increased LZ losses. The reasons for this are twofold. Firstly the smaller rings allow the atom cloud to sample a larger fraction of the resonant ellipsoid, including those regions where $\Omega_{R}(\textrm{\textbf{r}})\approx 0$. Secondly, increased trap frequencies result in the atoms traversing the avoided crossing with a higher relative velocity. Both factors combine to increase the probability of a diabatic transition. The lifetime for the smallest value of $r_{0} = \unit[50]{\mu \textrm{m}}$ is $<\unit[500]{\textrm{ms}}$. It is anticipated that by increasing rf power beyond that which is available in the current apparatus, the lifetime in this regime will be improved.

\begin{figure}[h]
\begin{center}
\includegraphics[width=1.0\columnwidth]{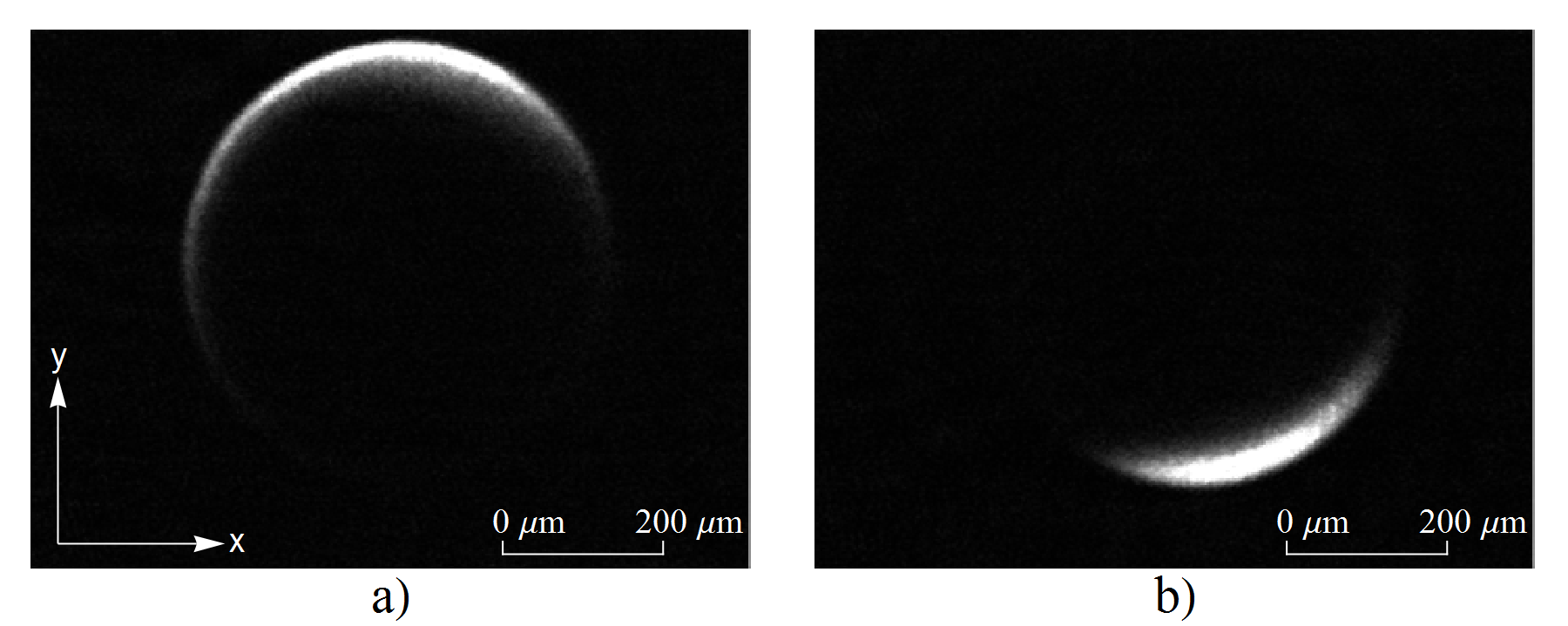}
\includegraphics[width=1.0\columnwidth]{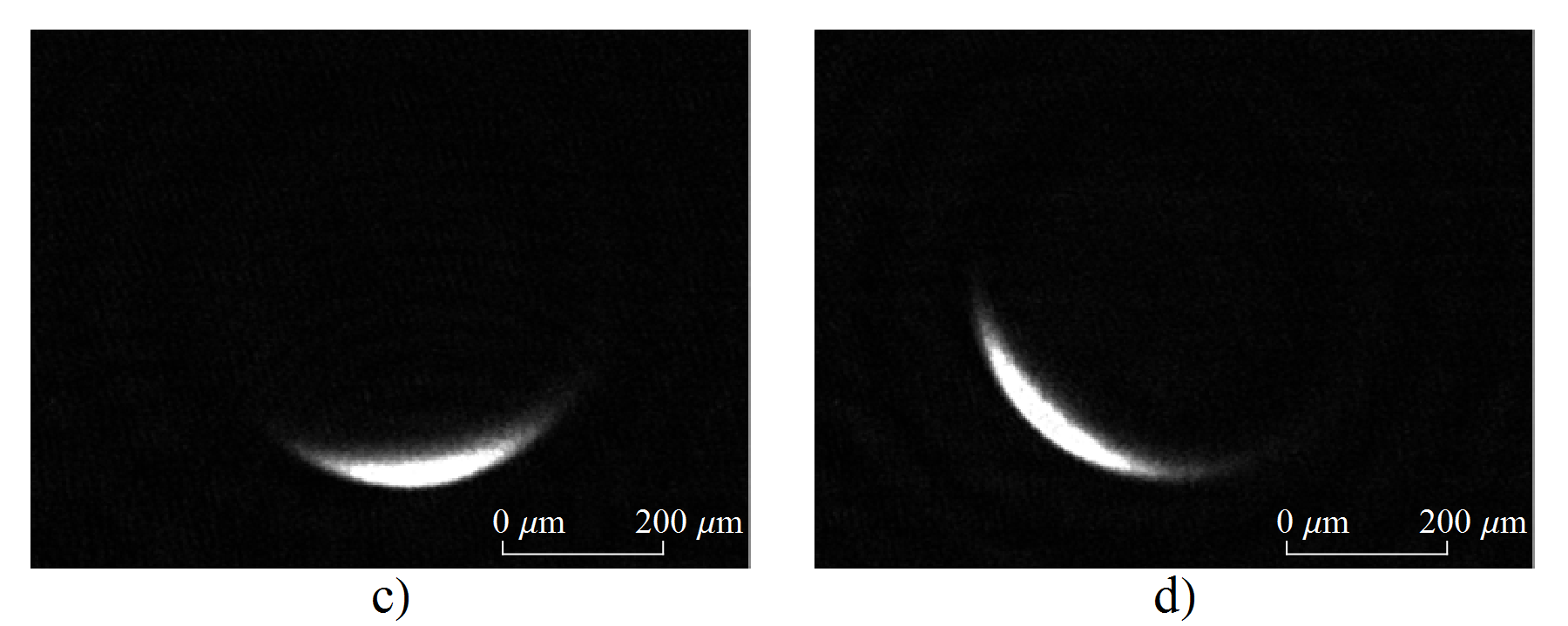}
\end{center}
\caption{\label{fig:ringtilt}Images of a BEC in a tilted ring trap. The direction of the tilt is controlled by the relative phase $\Delta_{z}$ of the axial dressing radiation $B^{z}_{rf}$. Here (a) $\Delta_{z}=0^{\circ}$, (b) $\Delta_{z}=90^{\circ}$, (c) $\Delta_{z}=180^{\circ}$ and (d) $\Delta_{z}=270^{\circ}$. A gravitational tilt in the ring trap prevents the atoms from responding linearly to changes in the rf polarisation. It is apparent that the adjustment is sufficient to offset the intrinsic tilt caused by imperfect alignment of the coils with respect to the vertical axis defined by gravity.}
\end{figure}
Misalignments of the quadrupole field symmetry axis with gravity resulting from the construction and positioning of the magnetic coils led to the first ring traps being tilted by approximately $2^{\circ}$. By exploiting the vectorial nature of $\Omega_{R}(\textrm{\textbf{r}})$ the balancing of the ring in the radial plane has been significantly improved such that a cloud at $\sim\unit[85]{\textrm{nK}}$ spreads fully around a ring of radius $r_{0}=\unit[238]{\mu \textrm{m}}$. The addition of an axially directed dressing field $\textbf{B}^{z}_{rf}(t)=B^{z}_{rf}\cos{(\omega_{rf}t+\Delta_{z})}\hat{\textrm{\textbf{e}}}_{z}$ tilts the rf polarisation vector. The tilting angle is determined by the magnitude of $\textbf{B}^{z}_{rf}$ whilst the relative phase $\Delta_{z}$ specifies the direction of the tilt (see figure \ref{fig:ringtilt}). Tilting the polarisation vector shifts the position of maximum coupling away from the South pole of the ellipsoid, allowing it to be used to compensate for inhomogeneities around the ring.
\section{Rotation}
\label{sec:rotation}
\begin{figure}[t]
\begin{center}
\includegraphics[width=0.98\columnwidth]{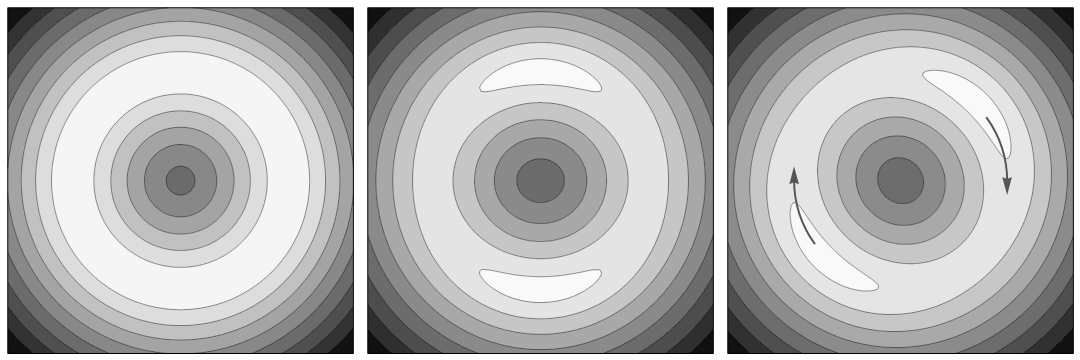}
\includegraphics[width=1.0\columnwidth]{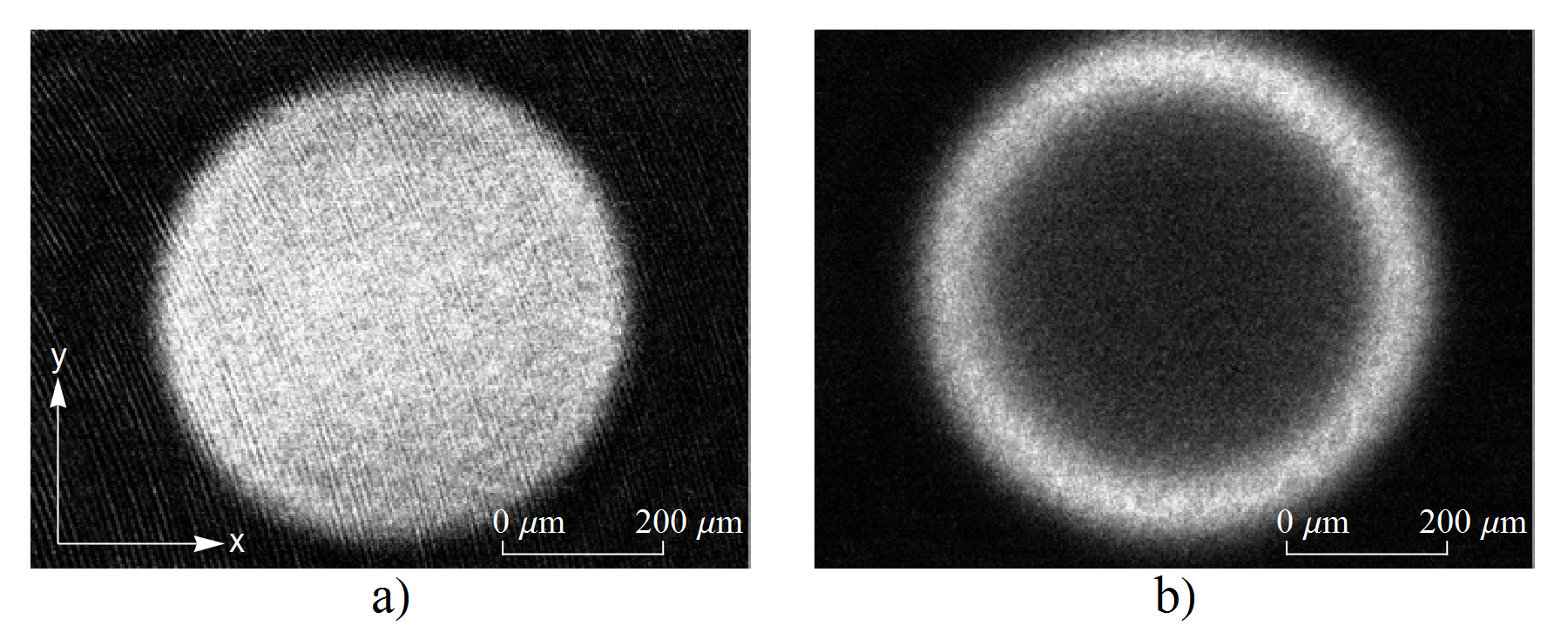}
\end{center}
\caption{\label{fig:rotation} Above: Contour plot of the rotation scheme in the $z=0$ plane. Circular symmetry around the ring is broken by transforming to elliptically polarised rf. Rotating the deformations acts to spin up the cloud. Below: Evidence for rotation. a) In the absence of rotation, when $B^{z}_{T}$ is ramped to zero the atoms return to the rf dressed shell potential where they are approximately uniformly distributed around the lower half of the ellipsoid (for the parameters listed below). b) For a rotating cloud (potential rotated at \unit[25]{Hz} for \unit[800]{ms}), the ring shape persists after the removal of the time-averaging field, because the atoms' angular momentum holds them at the equatorial regions of the ellipsoid. Both images are of a thermal cloud in a trap where $B'_{q} = \unit[71]{\textrm{G/cm}}$ and $B_{rf} = \unit[0.8]{\textrm{G}}$.}
\end{figure}

In order to study the dynamics of superflow in dilute atomic vapour BEC's trapped in ring potentials it is necessary to controllably introduce rotation into the system. In this experiment, as shown in figure \ref{fig:rotation}, a rotation scheme that utilises the vectorial nature of the coupling term $\Omega_{R}(\textrm{\textbf{r}})$ has successfully been employed to rotate atoms in the ring trap.
By relative adjustment of the dressing field amplitudes (such that $B^{x}_{rf} \neq B^{y}_{rf}$) the circularly symmetric coupling term takes on an elliptical form, introducing a spatially periodic variation around the ring. If the axis of this asymmetry is rotated, the atoms follow the variations in the potential and begin to orbit in the radial plane. To observe the dynamics of a rotating cloud, the circular symmetry of the potential is restored. Both processes of deformation and rotation are facilitated via a custom designed direct digital synthesis (DDS) frequency source, that employs user specified ramps to update the amplitude and phase of the dressing rf at sub-millisecond intervals.
\section{Conclusion and Outlook}
\label{sec:outlook}
In conclusion, the realisation of a versatile ring trap for ultracold atoms with dynamically adjustable radius has been presented. The ring is an example of a TAAP that forms via the combination of the processes of time-averaging and rf dressing of an axisymmetric quadrupole trap. A convenient loading scheme from a TOP trap is used to load a BEC into the ring trap. A method for tilting to correct for imbalances of the potential around the ring, and a rotation scheme for imparting angular momentum on the trapped atoms have been demonstrated. When operated as a circular waveguide the potential for matter-wave interferometry in this trap is enhanced by the effective magnetic insensitivity of the atoms (because they are trapped in a superposition of $m_{F}$ states that has zero net magnetic moment). In future experiments it will be possible to place a condensate in ring where $r_{0}\approx \unit[50] {\mu \textrm{m}}$ and $\omega_{r} = \omega_{z} \approx 2 \pi \times \unit[500]{\textrm{Hz}}$ such that the quantum coherence extends around the ring, allowing a study on the nature of superfluidity; this is particularly interesting in lower dimensional systems that can be created by working with smaller numbers of atoms. 

This work has been supported by the Engineering and Physical Sciences Research Council under Grant N0. EP/E010873/1.

\end{document}